\documentclass{jpp}
\usepackage{graphicx}
\usepackage{epstopdf, epsfig}
\usepackage{hyperref}
\usepackage{color}
\hypersetup{
   colorlinks=true,
   urlcolor=cyan,
   linkcolor=blue,
   citecolor=blue,
   filecolor=magenta,
   pdftitle={Sharelatex Example},
   pdfpagemode=FullScreen,
}

\shorttitle{Drifting bi-Kappa plasmas: dispersion and stability}
\shortauthor{R.~A. L\'{o}pez {\it et al.}}

\title{General dispersion properties of magnetized plasmas with drifting bi-Kappa distributions.\\
DIS-K: DIspersion Solver for Kappa plasmas}

\author{R. A. L\'{o}pez\aff{1}
  \corresp{\email{rlopez186@gmail.com}},
  S.~M. Shaaban\aff{2,3}
 \and M. Lazar\aff{4,5}}

\affiliation{
\aff{1}Departamento de F\'{\i}sica, Universidad de Santiago de Chile, Usach, 9170124 Santiago, Chile
\aff{2}Institute of Experimental and Applied Physics, University of Kiel, Leibnizstrasse 11, D-24118 Kiel, Germany
\aff{3}Theoretical Physics Research Group, Physics Department, Faculty of Science, Mansoura University, 35516, Mansoura, Egypt
\aff{4}Centre for mathematical Plasma Astrophysics, KU Leuven, Celestijnenlaan 200B, B-3001 Leuven, Belgium
\aff{5}Institut f\"ur Theoretische Physik, Lehrstuhl IV: Weltraum- und Astrophysik, Ruhr-Universit\"at Bochum, D-44780 Bochum, Germany
}

\begin{document}

\maketitle

\begin{abstract}
Space plasmas are known to be out of (local) thermodynamic equilibrium, as observations show direct or indirect evidences of non-thermal velocity distributions of plasma particles. Prominent are the anisotropies relative to the magnetic field, anisotropic temperatures, field-aligned beams or drifting populations, but also, the suprathermal populations enhancing the high-energy tails of the observed distributions.
Drifting bi-Kappa distribution functions can provide a good representation of these features and enable for a kinetic fundamental description of the dispersion and stability of these collision-poor plasmas, where particle-particle collisions are rare but wave-particle interactions appears to play a dominant role in the dynamic.
In the present paper we derive the full set of components of the dispersion tensor for magnetized plasma populations modeled by drifting bi-Kappa distributions. A new solver called DIS-K (DIspersion Solver for Kappa plasmas) is proposed to solve numerically the dispersion relations of high complexity. The solver is validated by comparing to the damped and unstable wave solutions obtained with other codes, operating in the limits of drifting Maxwellian and non-drifting Kappa models. These new theoretical tools enable more realistic characterizations, both analytical and numerical, of wave fluctuations and instabilities in complex kinetic configurations measured in-situ in space plasmas.
\end{abstract}

\section{Introduction} \label{sec:intro}

In collision-poor plasmas from heliosphere, in-situ measurements of the velocity distributions of particles reveal a variety of non-thermal features, such as suprathermal populations and anisotropies with respect to the magnetic field direction, mainly anisotropic temperatures and field-aligned beams~\citep{Maksimovic-etal-2005, Marsch2006a, Wilson-etal-2019a}. Physical mechanisms that can maintain these states of non-equilibrium are mainly triggered by the small-scale (kinetic) wave turbulence and fluctuations, which are also confirmed by the observations \citep{Alexandrova-etal-2013}. If resonant wave-particle interactions can preferentially energize plasma particles leading to kinetic anisotropies \citep{Isenberg-Vasquez-2019}, stochastic acceleration by plasma turbulence may generate more diffuse suprathermal populations enhancing the high-energy tails of velocity distributions \citep{Bian-etal-2014}. The same wave-particle  interactions are also responsible for the excitation of instabilities \citep{Wilson2013, Bowen-etal-2020}, or further relaxation of anisotropic particles under the diffusion effect of the enhanced fluctuations \citep{Bale-etal-2009, Tong-etal-2019}. 
%
%
These fluctuations are mainly powered by the solar outflows inducing large-scale perturbations and wave turbulence, which are transported by the expanding solar wind and decay nonlinearly toward smaller (kinetic) scales where a perpetual exchange of energy between plasma waves and particles ultimately takes place \citep[and references therein]{Leamon1999, Verscharen2019a}. Representative are not only the mechanisms of resonant dissipation (damping) of waves leading to the heating of particles, but also the instabilities that can transfer (free) energy from anisotropic and suprathermal particles to waves and fluctuations \citep{Lazar-etal-2019,Shaaban-etalAA19, Verscharen2019b, Shaaban2021}. 

Locally, small-scale fluctuations can be markedly enhanced by the kinetic instabilities, sometimes rather than a turbulent cascade \citep{Gary2016, Woodham-etal-2019, Tong-etal-2019}. However, in the absence of energetic events (e.g., flares or coronal mass ejections) the wave fluctuations measured in-situ in space plasmas have wide bandwidths and small amplitudes, such that, their properties and effects on particles can be addressed in the frame of linear and quasilinear (QL) theories \citep{Gary1993, Gary2016, Yoon2017c, yoon-2019}. However, if the fluctuations are a superposition of nonlinear structures or fluid-like turbulent eddies, then linear and quasilinear theory do not apply.
Revealing these linear and QL properties of waves and instabilities in the solar wind, already perceived as a truly natural laboratory, has therefore acquired a special motivation. To do that, we need to derive the dispersion and stability relations, and, implicitly, the dielectric tensor of the plasma system relying directly on the shape of underlying velocity distributions \citep{Stix1992}.
The observed velocity distributions of plasma particles combine kinetic anisotropies, relative to the magnetic field direction, e.g., anisotropic temperatures \citep{Kasper2003, Stverak-etal-2008} or field aligned beams \citep{Pilipp1987, Marsch2006a}, and suprathermal tails well reproduced by the Kappa distribution functions \citep{Pierrard2010, Lazar-etal-2017}. The picture can be clarified by pointing on the details in the velocity distributions. For instance, the electrons show a low-energy core, well reproduced by a bi-Maxwellian distribution function, a suprathermal halo and an electron strahl, both of them well described by bi-Kappa distribution functions with relative drifts parallel to the magnetic field \citep{Maksimovic-etal-2005, Lazar-etal-2017, Wilson-etal-2019a, Wilson-etal-2019b}. 
The relative drift between core and halo is in general modest \citep{Wilson-etal-2019a} allowing the incorporation of these two components by another bi-Kappa, which is nearly bi-Maxwellian at low energies but decreases as a power-law at high energies \citep{Vasyliunas1968,Lazar-etal-2017}. 
It is thus obvious that Kappa models, including anisotropic variants like bi-Kappa, with or without relative drifts, are widely invoked for their success in modeling the observed distributions, not only for electrons, but also for protons and heavier ions \citep{Collier-etal-1996,Pierrard2010, Mason-Gloeckler-2012}.

Despite these observational evidences, in the dispersion and stability analysis the velocity distributions are still reduced to idealized bi-Maxwellian representations \citep{Verscharen2019b,Lopez2020a,Shaaban-Lazar-2020,Shaaban2021MNRAS}, for which the the dielectric tensor is already explicitly derived \citep{Stix1992}. In the context of drifting bi-Kappa plasmas, the analysis so far is limited only to the propagation parallel to the magnetic field \citep{Shaaban-etal-2018,Shaaban-etal-2020}, because the complete expression of the dielectric tensor, and, implicitly, the general dispersion relation for an arbitrary direction of propagation, are not trivial, but rather complicated, and have not been derived yet. Therefore, numerical solvers capable to resolve the full spectrum of waves and instabilities of this configuration do not yet exist.
In the present paper we provide the full expression of the dielectric tensor and general dispersion relation for magnetized plasma particles described by drifting bi-Kappa distribution functions. An advanced analysis becomes thus possible to characterize, of a more realistic manner, the dispersion and stability of plasma populations revealed by the in-situ observations. For instance, the suprathermal halo, and strahl electrons carrying the main heat flux in the solar wind \citep{Lazar-etal-2020a}, whose evolution with increasing the heliospheric distance \citep{Hammond-etal-1996, Maksimovic-etal-2005} are expected to be controlled by the self-generated instabilities \citep{Verscharen2019b,Lopez2020a,Jeong2020,Micera2020a}.

Also reported here is a new generalized DIspersion Solver for Kappa-distributed plasmas (abbreviated DIS-K), which implements the new dispersion tensor and numerically resolves the dispersion and (in)stability properties for all directions of propagation with respect to the magnetic field. Chronologically speaking, the first numerical solvers were built for Maxwellian plasmas, such as WHAMP~\citep{Roennmark1982} and PLADAWAN~\citep{Vinas2000}, or the more recent variants, PLUME~\citep{Klein2015} and NDHS~\citep{Verscharen2018}. 
Progress has also been made for other models, such as the Kappa distribution, which introduces new analytical and numerical challenges. Of major importance were the detailed studies of the new dispersion function for Kappa distributions \citep{Summers1991,Mace1995}, on which most of the codes developed later are based. 
Initially, efforts were dedicated to unmagnetized plasmas or reduced configurations in magnetized plasmas, such as electrostatic approximation or parallel propagation \citep{Hellberg2002,Lazar2009a, Vinas2017}. Earlier approaches to describe perpendicular and  oblique propagation \citep{Summers1994, Cattaert2007, Basu2009} have more recently been complemented by rigorous mathematical calculations of the general dielectric tensor providing compact and closed forms of its components \citep{Liu2014,Gaelzer2016,Gaelzer2016a, Kim2017a}. A relatively recent numerical implementation of the dielectric tensor for non-drifting bi-Kappa plasmas is DHSARK~\citep{Astfalk2015}, a pioneer solver for studying kinetic electromagnetic instabilities~\citep{Astfalk2018}.
Other similar codes (not named yet) were also developed in the last years, in order to extend the spectral analysis in bi-Kappa plasmas \citep{Lopez2019, Lazar-etal-2019}, and pave the way for a more elaborated tool, as our new solver DIS-K. In this paper we present the first numerical results obtained with this solver, which is capable to resolve more complex dispersion relations for plasma populations with drifting bi-Kappa populations, and the full spectrum of stable or unstable modes, of any nature, e.g., electrostatic or electromagnetic, periodic or aperiodic.

Our paper is organized as follows: in Sec.~\ref{sec:theory} we start by introducing the general (linear) dispersion tensor, and the drifting bi-Kappa parameterization for a  plasma of electrons and ions.  Explicit expressions of the newly derived components of the dispersion tensor are presented in Sec.~\ref{sec:disp_tensor}. We also show the agreement with different limit cases, e.g., Maxwellian, and non-drifting bi-Kappa, from previous studies. In  Sec.~\ref{sec:results} we solve numerically the dispersion relation for some illustrative cases specific to these limits, which allows us to compare with the previous results and test our numerical solver. Finally, our results are summarized in Sec.~\ref{sec:conclusions}.

%

\section{Theoretical formalism} \label{sec:theory}

Space plasmas are subject to multiple sources of inhomogeneities and temporal variations, such as wave turbulence, the interaction of fast and slow streams, or even the solar wind expansion. However, to describe the small-scale wave fluctuations and instabilities we can assume these plasmas sufficiently homogeneous, especially because linear and quasilinear (QL) theories seem to explain quite well the kinetic properties of plasma particles revealed by the in-situ observations \citep{Kasper2003, Stverak-etal-2008, Verscharen2019a}.
These are waves and instabilities which mainly depend on the nature of particle velocity distributions, and their analysis requires a kinetic approach.    

\subsection{General dispersion relation} \label{sec:disp}

Without loss of generality we assume cartesian coordinates ($x,y,z$) with $z$-axis parallel to the magnetic field ${\bf B}$, and with the wave vector ${\bf k}$ in the ($x-z$) plane, such that 
\begin{equation}
  \mathbf{k}\,=\,k_\perp\,\mathbf{\hat{x}}+k_\parallel\,\mathbf{\hat{z}}
\end{equation}
where $\parallel, \perp$ are defined with respect to the magnetic field direction.
From Vlasov-Maxwell equations one can derive the general expression of the dispersion relation \citep{Stix1992}
\begin{equation}
{\bf \Lambda} \cdot {\bf E} = 0 \label{disp-rel}
\end{equation}
in terms of the electric field of the wave fluctuation ${\bf E}({\bf k},\omega)$ and the dispersion tensor $\Lambda$. For arbitrary (but still gyrotropic) velocity distribution functions $F_a (v_\perp, v_\parallel)$ of plasma species of sort $a$ (e.g., $a = e,p,i$ for electrons, protons and ions, respectively) the components of the dispersion read as follows
\begin{eqnarray}
  \Lambda_{ij}(\mathbf{k},\omega)&=&
  \delta_{ij}-\frac{c^2k^2}{\omega^2}\left(\delta_{ij}-\frac{k_ik_j}{k^2}\right)
  \nonumber\\
  &&+\sum_a\frac{\omega_{pa}^2}{\omega^2}\int
  d\mathbf{v}\sum_{n=-\infty}^\infty
  \frac{V_i^nV_{j}^{n*}}{\omega-k_\parallel v_\parallel-n\Omega_a}
  \left(\frac{\omega-k_\parallel v_\parallel}{v_\perp}\frac{\partial
    F_a}{\partial v_\perp}+k_\parallel\frac{\partial F_a}{\partial
    v_\parallel}\right)\nonumber\\
  &&+\mathbf{\hat{B}}_i\mathbf{\hat{B}}_j\sum_a\frac{\omega_{pa}^2}{\omega^2}
  \int d\mathbf{v}v_\parallel\left(\frac{\partial F_a}{\partial
    v_\parallel}-\frac{v_\parallel}{v_\perp}\frac{\partial
    F_a}{\partial v_\perp}\right)\,.
    \label{eq:tensor_general}
\end{eqnarray}
Here
\begin{equation}
  \mathbf{V}^n\,=\,v_\perp\frac{nJ_n(b)}{b}\,\mathbf{\hat{e}}_1-iv_\perp
  J_n'(b)\mathbf{\hat{e}}_2+v_\parallel
  J_n(b)\mathbf{\hat{e}}_3\,,\,\,\,\,\mathbf{\hat{B}}=B_0\mathbf{\hat{e}}_3\,,\,\,\,\,b=\frac{k_\perp v_\perp}{\Omega_a}\,,
\end{equation}
and $J_n(b)$ is the Bessel function with $J'_n(b)$ its first derivative, $i$ is the imaginary unit, $c$ is the speed of light, $\omega_{pa}=\sqrt{4\pi n_a/m_a}$ is the plasma frequency, $\Omega_a=q_aB_0/(m_ac)$ the gyrofrequency, $q_a$ the charge, $m_a$ the mass, and $n_a$ the number density of species $a$, respectively. 

\subsection{Drifting bi-Kappa distribution}

For magnetized plasma particles in space environments realistic models able to reproduce the main departures from thermal equilibrium, i.e., anisotropies and suprathermal populations, are drifting bi-Kappa velocity distribution functions
\begin{equation}
\label{eq:Fk}
  F_a(v_\perp,v_\parallel)\,= \frac{1}{\pi^{3/2}}
  \frac{1}{\alpha_{\perp a}^2\alpha_{\parallel a}}
  \frac{\Gamma(\kappa_a+1)}{\kappa_a^{3/2}\Gamma(\kappa_a-1/2)}
  \left[1+\frac{(v_\parallel-U_a)^2}{\kappa_a\alpha_{\parallel a}^2}
    +\frac{v_\perp^2}{\kappa_a\alpha_{\perp a}^2}\right]^{-\kappa_a-1}\,.
\end{equation}
$\kappa_a$ is the power law index, $\Gamma(x)$ is the Gamma function, $U_a$ is the drift speed, and the parameters $\alpha_{\parallel,\perp}$, known as the most probable speed~\citep{Vasyliunas1968},
\begin{equation}
  \alpha_{\parallel a}\,=\,\left(\frac{2k_BT_{\parallel a}}{m_a}\right)^{1/2}\,,
  \quad \alpha_{\perp a}\,=\,\left(\frac{2k_BT_{\perp a}}{m_a}\right)^{1/2}\,,
  \label{eq:thermal_speed}
\end{equation}
correspond to the thermal speeds of the Maxwellian limit that approximately describe the low-energy core out of the Kappa distribution~\citep{Lazar2015,Lazar2016}. Here $k_B$ is the Boltzmann's constant. The thermal speeds are related to the kinetic temperature through the second-order moment of the distribution,
\begin{eqnarray}
  T_{\parallel a}^{(\kappa)}&=&\int d\mathbf{v}(v_\parallel-U_a)^2
  F_a(v_\perp,v_\parallel)
  \,\,=\,\frac{m_a}{2k_B}\frac{2\kappa_a}{2\kappa_a-3}\,\alpha_{\parallel a}^2\,,\\
  T_{\perp a}^{(\kappa)}&=&\int d\mathbf{v}v_\perp^2F_a(v_\perp,v_\parallel)\,=\,\frac{m_a}{2k_B}\frac{2\kappa_a}{2\kappa_a-3}\,\alpha_{\perp a}^2\,,
\end{eqnarray}
requiring $\kappa_a>3/2$. 
These kinetic temperatures are greater than the corresponding temperatures of the Maxwellian limit, introduced in Eq.~(\ref{eq:thermal_speed}), through $T_{\parallel,\perp}=\lim_{\kappa\to\infty}T_{\parallel,\perp}^{(\kappa)}<T_{\parallel,\perp}^{(\kappa)}$.

\section{Dispersion tensor}\label{sec:disp_tensor}
We substitute the drifting bi-Kappa distribution function (\ref{eq:Fk}) in the general expression (\ref{eq:tensor_general}) of the dispersion tensor, and after integration obtain for each element of the dispersion tensor the following expressions
\begin{eqnarray}
  \Lambda_{ij}&=&\left(\begin{array}{ccc}
    D_{11} & iD_{12} & D_{13} \\
    -iD_{12} & D_{22} & iD_{23}\\
    D_{13} & -iD_{23} & D_{33}
  \end{array}\right)_{ij}\\
  \label{D11}
  D_{11} &=& 1-\frac{c^2k_\parallel^2}{\omega^2}
  +\sum_a\frac{\omega_{pa}^2}{\omega^2}
  \sum_{n=-\infty}^\infty \frac{n^2}{\lambda_a}
  \left[\xi_a Z_{n,\kappa}^{(1,2)}(\lambda_a,\zeta_a^n)
    +\frac{A_a}{2}\frac{\partial}{\partial\zeta_a^n}
    Z_{n,k}^{(1,1)}(\lambda_a,\zeta_a^n)\right]\,,\\
  D_{22} &=& 1-\frac{c^2k^2}{\omega^2}
  +\sum_a\frac{\omega_{pa}^2}{\omega^2}
  \sum_{n=-\infty}^\infty 
  \left[\xi_a W_{n,\kappa}^{(1,2)}(\lambda_a,\zeta_a^n)
    +\frac{A_a}{2}\frac{\partial}{\partial\zeta_a^n}
    W_{n,k}^{(1,1)}(\lambda_a,\zeta_a^n)\right]\,,\\
  D_{12} &=& \sum_a\frac{\omega_{pa}^2}{\omega^2}
  \sum_{n=-\infty}^\infty n
  \left[\xi_a \frac{\partial}{\partial \lambda_a}
    Z_{n,\kappa}^{(1,2)}(\lambda_a,\zeta_a^n)
    +\frac{A_a}{2}\frac{\partial^2}{\partial\lambda_a\partial\zeta_a^n}
    Z_{n,k}^{(1,1)}(\lambda_a,\zeta_a^n)\right]\,,\\
  D_{13} &=& \frac{c^2k_\perp k_\parallel}{\omega^2}
  +2\sum_a\frac{q_a}{|q_a|}\frac{\omega_{pa}^2}{\omega^2}
  \sqrt{\frac{T_{\parallel a}}{T_{\perp a}}}
  \frac{U_a}{\alpha_{\parallel a}}
  \sum_{n=-\infty}^\infty
  \frac{n}{\sqrt{2\lambda_a}}
  \xi_a Z_{n,\kappa}^{(1,2)}(\lambda_a,\zeta_a^n)
  \nonumber\\
  &&-\sum_a\frac{q_a}{|q_a|}\frac{\omega_{pa}^2}{\omega^2}
  \sqrt{\frac{T_{\parallel a}}{T_{\perp a}}}
  \sum_{n=-\infty}^\infty
  \frac{n}{\sqrt{2\lambda_a}}
  \left[\xi_a-A_a\left(\zeta_a^n+\frac{U_a}{\alpha_{\parallel a}}\right)\right]
  \frac{\partial}{\partial\zeta_a^n}Z^{(1,1)}_{n,\kappa}(\lambda_a,\zeta_a^n)
  \,,\\
  D_{23} &=& \sum_a\frac{\omega_{pa}^2}{\omega^2}\frac{|q_a|}{q_a}
  \sqrt{\frac{T_{\parallel a}}{T_{\perp a}}}
  \sqrt{\frac{\lambda_a}{2}}
  \sum_{n=-\infty}^\infty
    \left(\xi_a-A_a\left(\zeta_a^n+\frac{U_a}{\alpha_{\parallel a}}\right)\right)
    \frac{\partial^2}{\partial\lambda_a\partial\zeta_a^n}
    Z_{n,\kappa}^{(1,1)}(\lambda_a,\zeta_a^n)
  \nonumber\\
  &&-2\sum_a\frac{\omega_{pa}^2}{\omega^2}\frac{|q_a|}{q_a}
  \sqrt{\frac{T_{\parallel a}}{T_{\perp a}}}
  \sqrt{\frac{\lambda_a}{2}}
  \frac{U_a}{\alpha_{\parallel a}}
  \sum_{n=-\infty}^\infty
    \xi_a\frac{\partial}{\partial\lambda_a}
    Z_{n,\kappa}^{(1,2)}(\lambda_a,\zeta_a^n)
    \,,\\
  D_{33} &=&  1-\frac{c^2k_\perp^2}{\omega^2}
  +2\sum_a\frac{\omega_{pa}^2}{\omega^2}
  \frac{T_{\parallel a}}{T_{\perp a}}
  \left(1-\frac{1}{2\kappa_a}\right)\frac{U_a^2}{\alpha_{\parallel a}^2}
  +2\sum_a\frac{\omega_{pa}^2}{\omega^2}
  \frac{T_{\parallel a}}{T_{\perp a}}
  \frac{U_a^2}{\alpha_{\parallel a}^2}
  \sum_{n=-\infty}^\infty
  \xi_aZ_{n,\kappa}^{(1,2)}(\lambda_a,\zeta_a^n)
  \nonumber\\
  &&-\sum_a\frac{\omega_{pa}^2}{\omega^2}
  \frac{T_{\parallel a}}{T_{\perp a}}
  \sum_{n=-\infty}^\infty
  \left[(\xi_a-A_a\zeta_a^n)
    \left(\zeta_a^n+2\frac{U_a}{\alpha_{\parallel a}}\right)
    -A_a\frac{U_a^2}{\alpha_{\parallel a}^2}\right]
  \frac{\partial}{\partial\zeta_a^n}Z_{n,\kappa}^{(1,1)}(\lambda_a,\zeta_a^n)
  \,.
  \label{D33}
\end{eqnarray}
Here we have used the following definitions~\citep{Gaelzer2016,Gaelzer2016a,Kim2017a}
\begin{eqnarray}
  \label{eq:Znk}
  Z_{n,\kappa}^{(\alpha,\beta)}(\lambda,\xi) &=& 
  2\int_0^\infty dx\,
  \frac{xJ_n^2(x\sqrt{2\lambda})}{(1+x^2/\kappa)^{\kappa+\alpha+\beta-1}}
  Z_\kappa^{(\alpha,\beta)}
  \left(\frac{\xi}{\sqrt{1+x^2/\kappa}}\right)\,,\\
  \label{eq:Ynk}
  Y_{n,\kappa}^{(\alpha,\beta)}(\lambda,\xi) &=&
  \frac{2}{\lambda}\int_0^\infty dx\,
  \frac{x^3J_{n-1}(x\sqrt{2\lambda})
    J_{n+1}(x\sqrt{2\lambda})}{(1+x^2/\kappa)^{\kappa+\alpha+\beta-1}}
  Z_\kappa^{(\alpha,\beta)}
  \left(\frac{\xi}{\sqrt{1+x^2/\kappa}}\right)\,,\\
  \label{eq:Wnk}
  W_{n,\kappa}^{(\alpha,\beta)}(\lambda,\xi) &=&
  \frac{n^2}{\lambda}Z_{n,\kappa}^{(\alpha,\beta)}(\lambda,\xi)
  -2\lambda Y_{n,\kappa}^{(\alpha,\beta)}(\lambda,\xi)\,,\\
  \label{eq:Zk}
  Z_\kappa^{(\alpha,\beta)}(\xi) &=& \frac{1}{\pi^{1/2}\kappa^{\beta+1/2}}
  \frac{\Gamma(\kappa+\alpha+\beta-1)}{\Gamma(\kappa+\alpha-3/2)}
  \int_{-\infty}^\infty ds\,
  \frac{(1+s^2/\kappa)^{-(\kappa+\alpha+\beta-1)}}{s-\xi}\,,\\
  \lambda_a &=& \frac{k_\perp^2\alpha_{\perp a}^2}{2\Omega_a^2}\,,\\
  \xi_a &=& \frac{\omega-k_\parallel U_a}{k_\parallel\alpha_{\parallel a}}\,,\\
  \zeta_a^n&=&\frac{\omega-k_\parallel U_a-n\Omega_a}{k_\parallel\alpha_{\parallel
      a}}\,,\\
  A_a&=&1-\frac{T_{\perp a}}{T_\parallel a}\,.
\end{eqnarray}
A list of the specific expressions required in Eqs.~(\ref{D11})--(\ref{D33}) are provided in Appendix~\ref{appendix1} and \ref{appendix2}.\\
From Eq.~(\ref{disp-rel}), the dispersion relation for nontrivial solutions ($E \ne 0$) requires the determinant of the dispersion tensor to satisfy det$\{\Lambda\} = 0$, which can be written explicitly as
\begin{equation}
  0\,=\,D_{11}D_{22}D_{33}-D_{11}D_{23}^2-D_{22}D_{13}^2-D_{33}D_{12}^2-2D_{12}D_{13}D_{23}\,.
\end{equation}

In order to optimize the performance of the dispersion solver, we can further simplify the expressions of the dispersion tensor components in Eqs. (\ref{D11})--(\ref{D33}), to minimize the number of integrals that need to be performed. These components can be written as follows, 
\begin{eqnarray}
  D_{11} &=& 1-\frac{c^2k_\parallel^2}{\omega^2}
  +\sum_a\frac{\omega_{pa}^2}{\omega^2}
  \sum_{n=-\infty}^\infty \frac{n^2}{\lambda_a}
  I_{11}(\lambda_a,\zeta_a^n)\,,\\
  D_{22} &=& 1-\frac{c^2k^2}{\omega^2}
  +\sum_a\frac{\omega_{pa}^2}{\omega^2}
  \sum_{n=-\infty}^\infty 
  I_{22}(\lambda_a,\zeta_a^n)\,,\\
  D_{12} &=& \sum_a\frac{\omega_{pa}^2}{\omega^2}
  \sum_{n=-\infty}^\infty n
  I_{12}(\lambda_a,\zeta_a^n)\,,\\
  D_{13} &=& \frac{c^2k_\perp k_\parallel}{\omega^2}
  +\sum_a\frac{q_a}{|q_a|}\frac{\omega_{pa}^2}{\omega^2}
  \sqrt{\frac{T_{\parallel a}}{T_{\perp a}}}
  \sum_{n=-\infty}^\infty
  \frac{n}{\sqrt{2\lambda_a}}
  I_{13}(\lambda_a,\zeta_a^n)
  \,,\\
  D_{23} &=& \sum_a\frac{|q_a|}{q_a}\frac{\omega_{pa}^2}{\omega^2}
  \sqrt{\frac{T_{\parallel a}}{T_{\perp a}}}
  \sqrt{\frac{\lambda_a}{2}}
  \sum_{n=-\infty}^\infty
  I_{23}(\lambda_a,\zeta_a^n)
  \,,\\
  D_{33} &=&  1-\frac{c^2k_\perp^2}{\omega^2}
  +2\sum_a\frac{\omega_{pa}^2}{\omega^2}
  \frac{T_{\parallel a}}{T_{\perp a}}
  \left(1-\frac{1}{2\kappa_a}\right)\frac{U_a^2}{\alpha_{\parallel a}^2}
  +2\sum_a\frac{\omega_{pa}^2}{\omega^2}
  \frac{T_{\parallel a}}{T_{\perp a}}
  \sum_{n=-\infty}^\infty
  I_{33}(\lambda_a,\zeta_a^n)
  \,.
\end{eqnarray}
Now, each component of the dispersion tensor depends on a single integral of the form
\begin{eqnarray}
  \label{I11}
  I_{11}(\lambda_a,\zeta_a^n)&=&
  \int_0^\infty dx\,\frac{2xJ_n^2(x\sqrt{2\lambda_a})}{(1+x^2/\kappa_a)^{\kappa_a+3/2}}H_a^n(x)\,,\\
  I_{22}(\lambda_a,\zeta_a^n)&=&
  \int_0^\infty dx\,
  \frac{2x\left(\frac{n^2}{\lambda_a}J_n^2\left(x\sqrt{2\lambda_a}\right)-
  2x^2J_{n-1}\left(x\sqrt{2\lambda_a}\right)
  J_{n+1}\left(x\sqrt{2\lambda_a}\right)\right)}{(1+x^2/\kappa_a)^{\kappa+3/2}}
  H_a^n(x)\,,\\
  I_{12}(\lambda_a,\zeta_a^n) &=&
  \sqrt{\frac{2}{\lambda_a}}
  \int_0^\infty dx\,
  \frac{x^2J_n\left(x\sqrt{2\lambda_a}\right)\left[
  J_{n-1}\left(x\sqrt{2\lambda_a}\right)
  -J_{n+1}\left(x\sqrt{2\lambda_a}\right)\right]}{(1+x^2/\kappa)^{\kappa+3/2}}
  H_a^n(x)\,,\\
  I_{13}(\lambda_a,\zeta_a^n)&=&
  4\int_0^\infty
  dx\frac{xJ_n^2(x\sqrt{2\lambda_a})}{(1+x^2/\kappa_a)^{\kappa_a+3/2}}
  K_a^n(x)\,,\\
  I_{23}(\lambda_a,\zeta_a^n)&=&
  -\frac{4}{\sqrt{2\lambda_a}}\int_0^\infty dx
  \frac{x^2J_n\left(x\sqrt{2\lambda_a}\right)\left[
  J_{n-1}\left(x\sqrt{2\lambda_a}\right)
  -J_{n+1}\left(x\sqrt{2\lambda_a}\right)\right]}{(1+x^2/\kappa_a)^{\kappa_a+3/2}}
  K_a^n(x)\,,\\
  \label{I33}
  I_{33}(\lambda_a,\zeta_a^n)&=&  
  2\int_0^\infty dx
  \frac{xJ_n^2(x\sqrt{2\lambda_a})}{(1+x^2/\kappa_a)^{\kappa_a+3/2}}
  Q_a^n(x)\,,
  \end{eqnarray}
  with the following functions
  \begin{eqnarray}
   H_a^n(x) &=&-\left(1-\frac{1}{4\kappa_a^2}\right)A_a
   +\frac{(\xi_a-A_a\zeta_a^n)}{\sqrt{1+x^2/\kappa_a}}\,Z_\kappa^{(1,2)}
  \left(\frac{\zeta_a^n}{\sqrt{1+x^2/\kappa_a}}\right)
  \,,\\
  K_a^n(x)&=&\left(1-\frac{1}{4\kappa_a^2}\right)\xi_a+\left(\zeta_a^n+\frac{U_a}{\alpha_{\parallel a}}\right)H_a^n(x)
  \,,\\
  Q_a^n(x)&=&\xi_a\left(1-\frac{1}{4\kappa_a^2}\right)
  \left(\zeta_a^n+\frac{2U_a}{\alpha_{\parallel a}}\right)+\left(\zeta_a^n+\frac{U_a}{\alpha_{\parallel a}}\right)^2
  H_a^n(x)
    \,.
  \end{eqnarray}
In this way, we only need to implement the modified plasma dispersion function $Z_\kappa^{(1,2)}$. This function can be directly evaluated using Eq.~(\ref{Zk12-integer}) for integer values of $\kappa$, or Eq.~(\ref{F21}) for real values of $\kappa$. Integrals in Eqs.~(\ref{I11})--(\ref{I33}) are evaluated using an adaptive numerical quadrature based in the routine QUADPACK~\citep{Piessens1983}.\\

\subsection{Maxwellian limit}
\label{Maxwellian}

In the limit $\kappa_a\to\infty$ we recover the dispersion tensor for a drifting bi-Maxwellian plasma \citep{Stix1992}, by using the following limits
\begin{eqnarray}
  \lim_{\kappa_a\to\infty}Z_{n,k}^{(\alpha,\beta)}(\lambda_a,\zeta_a^n)&=&
  \Lambda_n(\lambda_a)Z(\zeta_a^n) 
  \\
  \lim_{\kappa_a\to\infty}Y_{n,k}^{(\alpha,\beta)}(\lambda_a,\zeta_a^n)&=&
  \Lambda'_n(\lambda_a)Z(\zeta_a^n) 
  \\
  \lim_{\kappa_a\to\infty}
  \left(1+y^2/\kappa_a\right)^{-(\kappa+\alpha)}&=&e^{-y^2} \,.
\end{eqnarray}
Here $\Lambda_n(x)=I_n(x)e^{-x}$, with $I_n(x)$ the modified Bessel function, and $Z(x)$ is the plasma dispersion function for Maxwellian distributed plasmas \citep{Fried1961}. Thus, the elements of the dispersion tensor reduce to
\begin{eqnarray}
  D_{11} &=& 1-\frac{c^2k_\parallel^2}{\omega^2}
  +\sum_a\frac{\omega_{pa}^2}{\omega^2}
  \sum_{n=-\infty}^\infty \frac{n^2}{\lambda_a}\Lambda_n(\lambda_a)\mathcal{A}_n
  \,,\\
  D_{22} &=& 1-\frac{c^2k^2}{\omega^2}
  +\sum_a\frac{\omega_{pa}^2}{\omega^2}
  \sum_{n=-\infty}^\infty 
  \left(\frac{n^2}{\lambda_a}\Lambda_n(\lambda_a)
  -2\lambda_a\Lambda'_n(\lambda_a)\right)\mathcal{A}_n
  \,,\\
  D_{12} &=& \sum_a\frac{\omega_{pa}^2}{\omega^2}
  \sum_{n=-\infty}^\infty n\Lambda_n(\lambda_a)\mathcal{A}_n
  \,,\\
  D_{13} &=&\frac{c^2k_\perp k_\parallel}{\omega^2}
  +2\sum_a\frac{q_a}{|q_a|}\frac{\omega_{pa}^2}{\omega^2}
  \sqrt{\frac{T_{\parallel a}}{T_{\perp a}}}
  \sum_{n=-\infty}^\infty
  \frac{n}{\sqrt{2\lambda_a}}
  \Lambda_n(\lambda_a)\mathcal{B}_n
  \,,\\
  D_{23} &=& -2\sum_a\frac{\omega_{pa}^2}{\omega^2}\frac{|q_a|}{q_a}
  \sqrt{\frac{T_{\parallel a}}{T_{\perp a}}}
  \sqrt{\frac{\lambda_a}{2}}
  \sum_{n=-\infty}^\infty
  \Lambda'_n(\lambda_a)\mathcal{B}_n
  \,,\\
  D_{33} &=& 1-\frac{c^2k_\perp^2}{\omega^2}
  +2\sum_a\frac{\omega_{pa}^2}{\omega^2}
  \frac{T_{\parallel a}}{T_{\perp a}}
  \frac{U_a}{\alpha_{\parallel a}}
  \left(\frac{U_a}{\alpha_{\parallel a}}+2\xi_a\right)
  +2\sum_a\frac{\omega_{pa}^2}{\omega^2}
  \frac{T_{\parallel a}}{T_{\perp a}}
  \sum_{n=-\infty}^\infty
  \Lambda_n(\lambda_a)\mathcal{C}_n
  \,,\\
  \mathcal{A}_n&=&-A_a+(\xi_a-A_a\zeta_a^n)Z(\zeta_a^n)
  \,,\\
  \mathcal{B}_n&=&\xi_a+\left(\zeta_a^n+\frac{U_a}{\alpha_{\parallel a}}\right)
    \mathcal{A}_n
  \\
  \mathcal{C}_n&=&\xi_a\zeta_a^n+\left(\zeta_a^n+\frac{U_a}{\alpha_{\parallel a}}\right)^2 \mathcal{A}_n \,.
\end{eqnarray}
These expressions are equivalent to those provided in \citet{Stix1992}, pp. 258--260.

\subsection{{Non-drifting bi-Kappa} }
\label{non-drifting}
For components (\ref{D11})--(\ref{D33}) of the dielectric tensor, in the nondrifting limit $U_a=0$ we recover the results for bi-Kappa plasmas \citep{Gaelzer2016a, Kim2017a}
\begin{eqnarray}
  D_{11} &=& 1-\frac{c^2k_\parallel^2}{\omega^2}
  +\sum_a\frac{\omega_{pa}^2}{\omega^2}
  \sum_{n=-\infty}^\infty \frac{n^2}{\lambda_a}
  \left[\xi_a Z_{n,\kappa}^{(1,2)}(\lambda_a,\zeta_a^n)
    +\frac{A_a}{2}\frac{\partial}{\partial\zeta_a^n}
    Z_{n,k}^{(1,1)}(\lambda_a,\zeta_a^n)\right]\,,\\
  D_{22} &=& 1-\frac{c^2k^2}{\omega^2}
  +\sum_a\frac{\omega_{pa}^2}{\omega^2}
  \sum_{n=-\infty}^\infty 
  \left[\xi_a W_{n,\kappa}^{(1,2)}(\lambda_a,\zeta_a^n)
    +\frac{A_a}{2}\frac{\partial}{\partial\zeta_a^n}
    W_{n,k}^{(1,1)}(\lambda_a,\zeta_a^n)\right]\,,\\
  D_{12} &=& \sum_a\frac{\omega_{pa}^2}{\omega^2}
  \sum_{n=-\infty}^\infty n
  \left[\xi_a \frac{\partial}{\partial \lambda_a}
    Z_{n,\kappa}^{(1,2)}(\lambda_a,\zeta_a^n)
    +\frac{A_a}{2}\frac{\partial^2}{\partial\lambda_a\partial\zeta_a^n}
    Z_{n,k}^{(1,1)}(\lambda_a,\zeta_a^n)\right]\,,\\
  D_{13} &=& \frac{c^2k_\perp k_\parallel}{\omega^2}
  -\sum_a\frac{q_a}{|q_a|}\frac{\omega_{pa}^2}{\omega^2}
  \sqrt{\frac{T_{\parallel a}}{T_{\perp a}}}
  \sum_{n=-\infty}^\infty
  \frac{n}{\sqrt{2\lambda_a}}
  \left(\xi_a-A_a\zeta_a^n\right)
  \frac{\partial}{\partial\zeta_a^n}Z^{(1,1)}_{n,\kappa}(\lambda_a,\zeta_a^n)
  \,,\\
  D_{23} &=& \sum_a\frac{\omega_{pa}^2}{\omega^2}\frac{|q_a|}{q_a}
  \sqrt{\frac{T_{\parallel a}}{T_{\perp a}}}
  \sqrt{\frac{\lambda_a}{2}}
  \sum_{n=-\infty}^\infty
    \left(\xi_a-A_a\zeta_a^n\right)
    \frac{\partial^2}{\partial\lambda_a\partial\zeta_a^n}
    Z_{n,\kappa}^{(1,1)}(\lambda_a,\zeta_a^n)
    \,,\\
  D_{33} &=&  1-\frac{c^2k_\perp^2}{\omega^2}
  -\sum_a\frac{\omega_{pa}^2}{\omega^2}
  \frac{T_{\parallel a}}{T_{\perp a}}
  \sum_{n=-\infty}^\infty
  \zeta_a^n(\xi_a-A_a\zeta_a^n)
  \frac{\partial}{\partial\zeta_a^n}Z_{n,\kappa}^{(1,1)}(\lambda_a,\zeta_a^n)
  \,.
\end{eqnarray}


\section{Numerical Results} \label{sec:results}
In this section we present the results obtained with our new solver DIS-K (DIspersion Solver for Kappa plasmas), which implements the new dispersion tensor for plasmas with drifting bi-Kappa distributions. This solver can also be used for any of the bi-Maxwellian or nondrifting limits discussed above.

    \begin{figure}
  \begin{center}
      \includegraphics[width=0.7\textwidth]{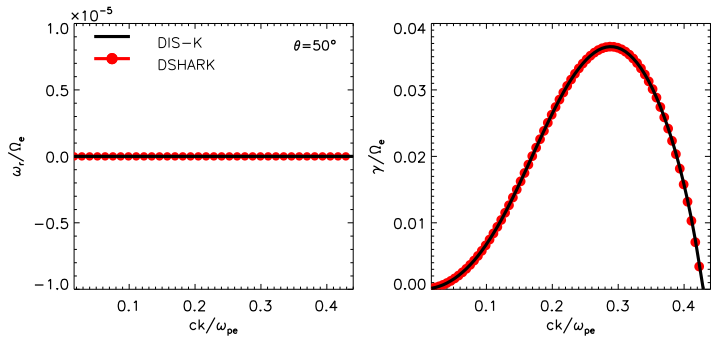}
      \caption{\label{disp_comparison_dshark} Unstable solutions of the aperodic electron firehose instability at $\theta=50^\circ$. In black are the solutions obtained with the new code, DIS-K, while red dots are obtained with DSHARK. Left panel shows the normalized (real) frequency, $\omega_r/\Omega_e$ vs. the normalized wave number $ck/\omega_{pe}$, and right panel shows the normalized growth rate, $\gamma/\Omega_e$ vs. $ck/\omega_{pe}$.}
  \end{center}
  \end{figure}

Here we show a number of illustrative examples of stable and unstable solutions obtained with the new solver DIS-K. As this is the first reported solver for drifting bi-Kappa plasmas, for its testing we can choose only from the limit configurations discussed above, and which can also be resolved by other solvers. As we know, DHSARK~\citep{Astfalk2015} is programmed to derive the unstable electromagnetic solutions of anisotropic plasma populations described by bi-Kappa distributions~\citep{Astfalk2018}. We will use the aperiodic electron firehose instability, as described in \citet{Shaaban2018a} and \citet{Lopez2019}, as a first test case. We consider an anisotropic electron distribution with $T_{\perp e}/T_{\parallel e}=0.5$, with $\beta_{\parallel e}=8\pi n_e T_{\parallel e}/B_0^2=4.0$, $\omega_{pe}/\Omega_e=100$, $\theta=50^\circ$, $\kappa_e=4$, and protons are modeled using an isotropic Maxwellian distribution with $\beta_p=4.0$. Fig.~\ref{disp_comparison_dshark} displays the unstable solutions obtained solving the set of equations presented in section 3.2. Left panel shows the real part of the normalized frequency, $\omega_r/\Omega_e$ vs. the normalized wave number $ck/\omega_{pe}$. As expected from the aperiodic nature of this instability, the wave frequency is zero for the entire range of unstable modes. Right panel of this figure shows the growth rate, $\gamma/\Omega_e$ vs. $ck/\omega_{pe}$, which coincides with that obtained using DSHARK in \citet{Shaaban2018a}, plotted here with red dots. The solutions of both codes show a perfect match for this case, validating our new solver DIS-K in the non-drifting bi-Kappa limit, $U_a=0$, discussed in Sec.~\ref{non-drifting}. 

\begin{figure}
  \begin{center}
      \includegraphics[width=0.7\textwidth]{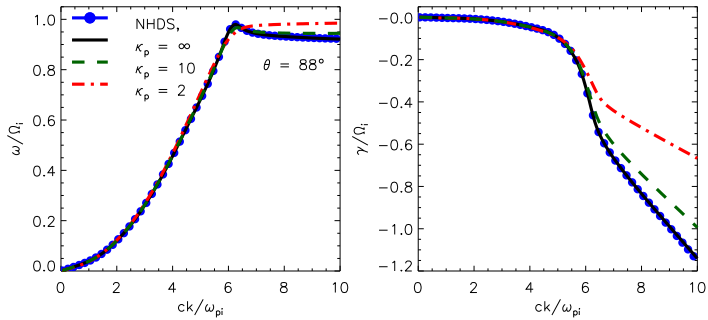}
      \caption{\label{disp_comparison_damped}The damped KAW solutions at $\theta=88^\circ$: blue dots are obtained with NHDS, and those obtained with the present code, DIS-K, are black lines for $\kappa_p=\infty$ (Maxwellian limit), green dot-dashed line for $\kappa_p=10$, and red dotted line for $\kappa_p=2$. Left panel shows $\omega_r/\Omega_i$ vs. $ck/\omega_{pi}$, and right panel  $\gamma/\Omega_i$ vs. $ck/\omega_{pi}$.}
  \end{center}
  \end{figure}

We now discuss damped solutions in order to test the analytical continuation of our modified plasma dispersion function. Expression in Eq.~(\ref{F21}) for arbitrary $\kappa$ or Eq.~(\ref{Zk12-integer}) for integer $\kappa$, already satisfy the Landau prescription, being analytically continuous through the entire complex frequency plane~\citep{Summers1991,Mace1995,Gaelzer2016}. 
In this case we study a damped Kinetic Alfv\'en Wave (KAW) propagating at highly oblique angles, $\theta=88^\circ$. We use $\beta_e=\beta_p=2.0$, $v_A/c=2.33\times10^{-4}$ (or $\omega_{pe}/\Omega_{ce}=100)$, where $v_A=B_0/\sqrt{4\pi n_pm_p}$ is the Alfv\'en speed. In Fig.~\ref{disp_comparison_damped} we compare our results with those obtained with another solver, NHDS~\citep{Verscharen2018} providing solutions in the Maxwellian limit. Blue dots correspond to the solution obtained with NHDS, while black line is our solution in the Maxwellian limit, $\kappa_p=\infty$. We have also included solutions for $\kappa_p=10$ (green dot-dashed lines) and $\kappa_p=2$ (red dotted lines), to show the behaviour of our code in the Kappa regime. In the Maxwellian limit both codes show a perfect agreement. For finite but large values of $\kappa_p$ the solution is similar to the Maxwellian case, but differences start appearing at large wave numbers. Overall, suprathermal particles reduce the damping of KAWs at large wave numbers~\citep{Kim2018}.
 
\begin{figure}
  \begin{center}
      \includegraphics[width=0.7\textwidth]{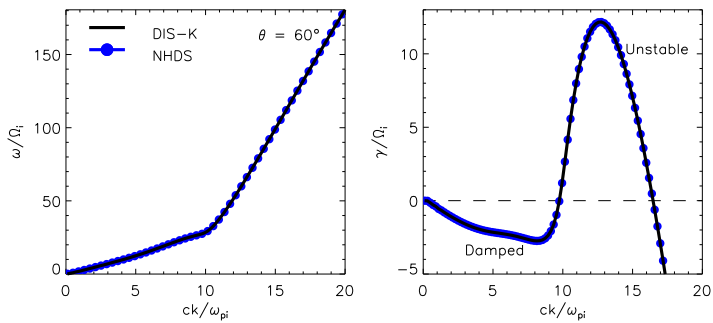}
      \caption{\label{disp_comparison}The oblique whistler heat-flux instability at $\theta=60^\circ$: black lines are solutions obtained with the present code, DIS-K, while blue dots correspond to NHDS. Left panel shows $\omega_r/\Omega_i$ vs. $ck/\omega_{pi}$, and right panel  $\gamma/\Omega_i$ vs. $ck/\omega_{pi}$.}
  \end{center}
  \end{figure}

Following, we test the drifting case. DSHARK is not programmed to solve the dispersion relation for drifting bi-Kappa distributions, and NHDS solver is only foreseen to operate with drifting bi-Maxwellian distributions, as described in Sec.~\ref{Maxwellian}. Therefore, we have to settle for this limiting case.
For this comparison we choose conditions for the oblique whistler heat-flux instability, as described in \citet{Lopez2020a}. We consider a plasma composed by two electron populations, a dense central core (subscript c) and a tenuous suprathermal beam (subscript b), with a relative drift along the background magnetic field, with the following properties: $n_c/n_0=0.95$, $n_b/n_0=0.05$ are the core and beam normalized number density, respectively, $T_{\parallel b}/T_{\parallel c}=4.0$, $T_{\perp j}/T_{\parallel j}=1.0$, $\omega_{pe}/\Omega_e=100$, $\beta_c=8\pi n_0T_c/B_0^2=2.0$ and $U_b/c=0.035$ (or $U_b/v_A=150$).
  Fig.~\ref{disp_comparison} shows the dispersion relation obtained for $\theta=60^\circ$. This time we plot the solution obtained using NHDS solver as a blue dots. In the range of wave numbers resolved we show both damped and unstable modes. We can clearly observe the agreement between both codes, in the real and imaginary part of the frequency, for damped and unstable modes.

\begin{figure}
  \begin{center}
      \includegraphics[width=0.95\textwidth]{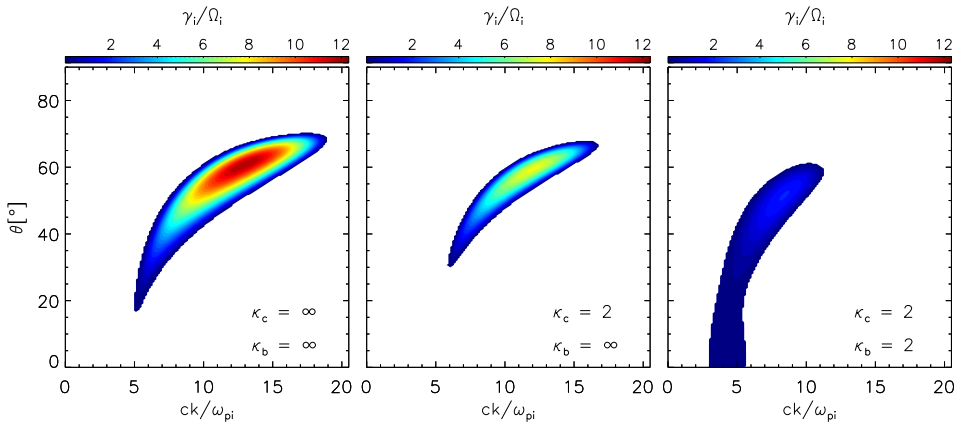}
      \caption{\label{maps}The oblique whistler heat-flux growth rates for all angles of propagation, and for various core-beam configurations described in the text.}
  \end{center}
  \end{figure}

Finally, assuming the same plasma conditions, we explore the same instability under the influence of suprathermal electrons, showing the results obtained with DIS-K for drifting bi-Kappa electrons. Fig.~\ref{maps} shows comparatively three different cases. For comparison, left panel shows the solution for drifting Maxwellian electrons, as in Fig.~\ref{disp_comparison}, but this time for the entire range of angles of propagation. Middle panel displays a new case, when core electrons are modeled by a Kappa distribution with $\kappa_c=2$, while the beam remains Maxwellian ($\kappa_b=\infty$). Right panel shows the case when both, core and beam populations are modeled by Kappa distribution with the same Kappa index, $\kappa_c=\kappa_b=2$. We observe that suprathermal electrons suppress the oblique whistler heat-flux instability. When the core is Kappa distributed, the range of unstable angles and wavenumbers is reduced, as well as the level of the unstable modes. The suppressing effect is even more prominent when both populations are Kappa distributed, reaching lower growth rates and shifting to lower angles. We can identify some possible explanations of these inhibiting effects, either in  the core damping, which counter-acts the strahl-driving of the oblique whistler instability, but also by a change in resonance conditions (populations involved, etc.), in this case, implying both Landau and cyclotron resonances.
On the other hand, the enhanced suprathermal tails reduce the effective excess of free energy in parallel (drifting) direction (with increasing the tails the core and beam electrons tend to merge and combine with each other, reducing the relative drift between them). Indeed, in the last case, we also obtain unstable modes in parallel and quasi-parallel (low angles) directions, specific to lower drifts.
  
\section{Conclusions}
\label{sec:conclusions}

We have presented the full set of components of the dielectric tensor for magnetized plasma populations modeled by drifting bi-Kappa distributions. This extended dielectric tensor has been implemented in a new dispersion solver, named DIS-K, and capable to resolve the full spectra of stable and unstable modes of these complex plasma distributions. In order to validate our results and our code, we carried out  illustrative cases enabling comparison with limiting conditions, e.g., nondrifting bi-Kappa and drifting bi-Maxwellian plasmas, resolved by the existing solvers. For comparison, the  the aperiodic electron firehose solutions driven by (nondrifting) bi-Kappa electrons are obtained with DSHARK, and we found a perfect agreement for both the wave-number dispersion of the wave frequency and growth rate. The same remarkable agreement has been obtained with the stable and unstable whistler like modes triggered by drifting bi-Maxwellian electrons populations and described by another solver, NHDS. We have also shown the capabilities of our code to handle damped solutions, as shown for the KAW dispersion curves, showing a perfect agreement when compared with NHDS in the Maxwellian limit. Further, we have explored the influence of suprathermal electrons on the oblique whistler heat-flux instability, showing that the instability is inhibited, i.e., growth rates are diminished (and the range of unstable wave-numbers is reduced), when core and beam electrons are modeled by drifting bi-Kappa distributions. We plan make this new code, DIS-K, publicly available in the near future.

These are new theoretical and numerical tools, which extend and improve the existing capabilities of analysis of the wave fluctuations, stable or unstable modes, specific to the complex particles configurations unveiled by the in-situ observations in space plasmas. Thus, applications can be considered in the context of solar wind and planetary environments, where all plasma species (electrons and ions) exhibit multiple drifting components, namely, core, halo and strahl, populated by suprathermals and with intrinsic anisotropies (i.e., anisotropic temperatures).\\ 

{\bf Acknowlegements}\\
We thank J.~A. Araneda for his valuable contribution to the development of this code. We also thank the anonymous reviewers for their insightful comments. R.A.L acknowledges the support of ANID Chile through FONDECyT grant No.~11201048. We also acknowledge the support of the projects SCHL 201/35-1 (DFG-German Research Foundation), C14/~19/089 (C1 project Internal Funds KU Leuven), G.0A23.16N (FWO-Vlaanderen), and C~90347 (ESA Prodex). S.M. Shaaban acknowledges the Alexander-von-Humboldt Research Fellowship, Germany.

\appendix

\section{Summary of necessary functions}
\label{appendix1}

\begin{eqnarray}
  Z_{n,\kappa}^{(1,2)}(\lambda_a,\zeta_a^n) \,&=&
  2\int_0^\infty dx\,
  \frac{xJ_n^2(x\sqrt{2\lambda_a})}{(1+x^2/\kappa)^{\kappa+2}}
  Z_\kappa^{(1,2)}
  \left(\frac{\zeta_a^n}{\sqrt{1+x^2/\kappa}}\right)\,,\\
  \frac{\partial}{\partial\zeta_a^n}
  Z_{n,k}^{(1,1)}(\lambda_a,\zeta_a^n) \,&=&
  2\int_0^\infty dx\,
  \frac{xJ_n^2(x\sqrt{2\lambda_a})}{(1+x^2/\kappa)^{\kappa+3/2}}
  {Z'}_\kappa^{(1,1)}
  \left(\frac{\zeta_a^n}{\sqrt{1+x^2/\kappa}}\right)\,,\\
  Y_{n,\kappa}^{(1,2)}(\lambda_a,\zeta_a^n) \,&=&
  \frac{2}{\lambda_a}\int_0^\infty dx\,
  \frac{x^3J_{n-1}(x\sqrt{2\lambda_a})
    J_{n+1}(x\sqrt{2\lambda_a})}{(1+x^2/\kappa)^{\kappa+2}}
  Z_\kappa^{(1,2)}
  \left(\frac{\zeta_a^n}{\sqrt{1+x^2/\kappa}}\right)\,,\\
  W_{n,\kappa}^{(1,2)}(\lambda_a,\zeta_a^n) \,&=&
  \frac{n^2}{\lambda_a}Z_{n,\kappa}^{(1,2)}(\lambda_a,\zeta_a^n)
  -2\lambda_a Y_{n,\kappa}^{(1,2)}(\lambda_a,\zeta_a^n)\,,\\
  \frac{\partial}{\partial\zeta_a^n}
  Y_{n,k}^{(1,1)}(\lambda_a,\zeta_a^n) \,&=&
  \frac{2}{\lambda_a}\int_0^\infty dx\,
  \frac{x^3J_{n-1}(x\sqrt{2\lambda_a})J_{n+1}(x\sqrt{2\lambda_a})}
       {(1+x^2/\kappa_a)^{\kappa_a+3/2}}
  {Z'}_\kappa^{(1,1)}
  \left(\frac{\zeta_a^n}{\sqrt{1+x^2/\kappa}}\right)\,,\\
  \frac{\partial}{\partial\zeta_a^n}
  W_{n,k}^{(1,1)}(\lambda_a,\zeta_a^n) \,&=&
  \frac{n^2}{\lambda_a}
  \frac{\partial}{\partial\zeta_a^n}Z_{n,\kappa}^{(1,1)}(\lambda_a,\zeta_a^n)
  -2\lambda_a
  \frac{\partial}{\partial\zeta_a^n}Y_{n,\kappa}^{(1,1)}(\lambda_a,\zeta_a^n)\,,\\
  \frac{\partial}{\partial\lambda_a}
  Z_{n,\kappa}^{(1,2)}(\lambda,\zeta_a^n) &=&
  \frac{2}{\sqrt{2\lambda_a}}\int_0^\infty dx\,
  \frac{x^2J_n(x\sqrt{2\lambda_a})[J_{n-1}(x\sqrt{2\lambda_a})-J_{n+1}(x\sqrt{2\lambda_a})]}
       {(1+x^2/\kappa)^{\kappa+2}}
       \nonumber\\
  &&\times Z_\kappa^{(1,2)}
  \left(\frac{\zeta_a^n}{\sqrt{1+x^2/\kappa}}\right)\,,\\
  \frac{\partial^2}{\partial\lambda\partial\zeta_a^n}
  Z_{n,k}^{(1,1)}(\lambda_a,\zeta_a^n) &=&
  \frac{2}{\sqrt{2\lambda_a}}\int_0^\infty dx\,
  \frac{x^2J_n(x\sqrt{2\lambda_a})[J_{n-1}(x\sqrt{2\lambda_a})-J_{n+1}(x\sqrt{2\lambda_a})]}
       {(1+x^2/\kappa)^{\kappa+3/2}}
       \nonumber\\
  &&\times{Z'}_\kappa^{(1,1)}
  \left(\frac{\zeta_a^n}{\sqrt{1+x^2/\kappa}}\right)\,.
\end{eqnarray}

\section{Some Useful Functions}
\label{appendix2}
The modified $Z_{\kappa}^{(\alpha,\beta)}$ function can be calculated in terms of the hypergeometric function as \citep{Gaelzer2016}
\begin{eqnarray}
  Z_\kappa^{(\alpha,\beta)}(\zeta_a^n)&=&
  \frac{i}{\kappa^{\beta+1}}
  \frac{\Gamma(\kappa+\alpha+\beta-1)\Gamma(\kappa+\alpha+\beta-1/2)}{\Gamma(\kappa)\Gamma(\kappa+\alpha-3/2)}
  \nonumber\\
  &&
  \times\,_2F_1\left[1,2(\kappa+\alpha+\beta-1);\lambda;\left(\frac{i\sqrt{\kappa}-\zeta_a^n}{2i\sqrt{\kappa}}\right)\right]\,,
\label{F21}
\end{eqnarray}
then, using
\begin{equation}
\frac{d}{dz}\,_2F_1[a,b;c;z]\,=\,\frac{ab}{c}\,_2F_1[a+1,b+1;c+1;z]\,,
\end{equation}
we have
\begin{eqnarray}
  \frac{\partial}{\partial\zeta_a^n}
  Z_\kappa^{(\alpha,\beta)}(\zeta_a^n) \,&=& 
  -\frac{1}{\kappa^{\beta+1}}
  \frac{\Gamma(\kappa+\alpha+\beta)\Gamma(\kappa+\alpha+\beta-1/2)}
       {\Gamma(\kappa+\alpha+\beta+1)\Gamma(\kappa+\alpha-3/2)}
       \nonumber\\
       &&\times
       \,_2F_1\left[2,2(\kappa+\alpha+\beta-1)+1;\kappa+\alpha+\beta+1;
         \left(\frac{i\sqrt{\kappa}-\zeta_a^n}{2i\sqrt{\kappa}}\right)\right]\,,
\end{eqnarray}

Also, a usefull expression is obtained from \citet{Gaelzer2016},
\begin{eqnarray}
  {Z'}^{(\alpha,\beta)}_\kappa(\zeta_a^n)&=&
  -2\left[\frac{\Gamma(\kappa+\alpha+\beta-1/2)}
    {\kappa^{\beta+1}\Gamma(\kappa+\alpha-3/2)}
    +\zeta_a^nZ_\kappa^{(\alpha,\beta+1)}(\zeta_a^n)\right]
\end{eqnarray}

We can obtain a simpler expression if $\kappa$ is assumed
integer, as in \citet{Summers1991}. Then we have
\begin{eqnarray}
  Z_\kappa^{(\alpha,\beta)}(\xi)&=&\frac{i}{2^{2(1-\lambda)}\kappa^{\beta+1/2}}
  \frac{\Gamma(\lambda-1)^2\Gamma(\lambda-1/2)}{\Gamma(\kappa+\alpha-3/2)\Gamma[2(\lambda-1)]}
  \left(\frac{\kappa+\xi^2}{\kappa}\right)^{1-\lambda}
  \nonumber\\
  &&\times\left\{1
  -\left(\frac{i\sqrt{\kappa}+\xi}{2i\sqrt{\kappa}}\right)^{\lambda-1}
  \frac{1}{\Gamma(\lambda-1)}\sum_{\ell=0}^{\lambda-2}
  \frac{\Gamma[\ell+\lambda-1]}{\Gamma(\ell+1)}
  \left(\frac{i\sqrt{\kappa}-\xi}{2i\sqrt{\kappa}}\right)^{\ell}
  \right\}
  \,. 
  \label{eq:Zint}
\end{eqnarray}

Let's take a look to some particular values
\begin{eqnarray}
  Z_\kappa^{(1,1)}(\xi)&=&\frac{2i\pi^{1/2}}{\kappa^{3/2}}
  \frac{\kappa!}{\Gamma(\kappa-1/2)}
  \left(\frac{\kappa+\xi^2}{\kappa}\right)^{-\kappa-1}
  \nonumber\\
  &&\left\{1
  -\frac{1}{\kappa!}\left(\frac{i\sqrt{\kappa}+\xi}{2i\sqrt{\kappa}}\right)^{\kappa+1}
  \sum_{\ell=0}^{\kappa}
  \frac{(\ell+\kappa)!}{\ell!}
  \left(\frac{i\sqrt{\kappa}-\xi}{2i\sqrt{\kappa}}\right)^{\ell}
  \right\}\,.
\end{eqnarray}

And we are also interested in
\begin{eqnarray}
  Z_\kappa^{(1,2)}(\xi)&=&\frac{2i\pi^{1/2}}{\kappa^{5/2}}
  \frac{(\kappa+1)!}{\Gamma(\kappa-1/2)}
  \left(\frac{\kappa+\xi^2}{\kappa}\right)^{-\kappa-2}
  \nonumber\\
  &&
  \left\{1
  -\frac{1}{(\kappa+1)!}\left(\frac{i\sqrt{\kappa}+\xi}{2i\sqrt{\kappa}}\right)^{\kappa+2}
  \,\sum_{\ell=0}^{\kappa+1}
  \frac{(\ell+\kappa+1)!}{\ell!}
  \left(\frac{i\sqrt{\kappa}-\xi}{2i\sqrt{\kappa}}\right)^{\ell}
  \right\}\,.
  \label{Zk12-integer}
\end{eqnarray}

For the derivative, we have
\begin{eqnarray}
  {Z'}_\kappa^{(1,1)}(\zeta_a^n)  &=&
  -\frac{i\zeta_a^n}{\kappa^{5/2}}
  \frac{\Gamma(\kappa+2)^2\Gamma(\kappa+3/2)}{\Gamma(\kappa-1/2)\Gamma(2\kappa+3)}
  \left(\frac{\kappa+(\zeta_a^n)^2}{4\kappa}\right)^{-\kappa-2}
\left\{1
  -\frac{\sqrt{\kappa}}{i\zeta_a^n}
  \left(\frac{i\sqrt{\kappa}+\zeta_a^n}{2i\sqrt{\kappa}}\right)^{\kappa+2}
  \right.
    \nonumber\\
  &&\left.\times
  \frac{1}{\Gamma(\kappa+2)}
  \sum_{\ell=0}^{\kappa+1}
  \frac{\Gamma(\ell+\kappa+1)}{\Gamma(\ell+1)}
  (\ell-\kappa-1)
  \left(\frac{i\sqrt{\kappa}-\zeta_a^n}{2i\sqrt{\kappa}}\right)^{\ell}
  \right\}\,.
\end{eqnarray}



\end{document}